\documentclass[a4paper, twocolumn]{article}
\usepackage[total={180mm, 250mm}]{geometry}
\usepackage{graphicx}

\title{\textbf{Spectral maximum in the terahertz photoconductance\\of a quantum point contact}}

\author{D.\,M.\,Kazantsev$^{1,2}$,
V.\,L.\,Alperovich$^{1,2,}\footnote{e-mail: alper@isp.nsc.ru}$,
V.\,A.\,Tkachenko$^{1,2,3}$,
Z.\,D.\,Kvon$^{1,2}$\\
$^1${\small\it Rzhanov Institute of Semiconductor Physics, 630090, Novosibirsk, Russia}\\
$^2${\small\it Novosibirsk State University, 630090, Novosibirsk, Russia}\\
$^3${\small\it Novosibirsk State Technical University, 630073, Novosibirsk, Russia}}

\date{\small August 5, 2022}

\begin{document}

\twocolumn[\maketitle\vspace{-0.5em}\centering\begin{minipage}{.8\textwidth}\setlength{\parindent}{1em}The disappearance of the giant terahertz photoconductance of a quantum point contact under the increase in the photon energy, which was discovered experimentally (Otteneder et al., Phys. Rev. Applied 10 (2018) 014015) and studied by the numerical calculations of the photon-stimulated transport (O.A. Tkachenko et al., JETP Lett. 108 (2018) 396), is explained by the momentum conservation upon absorption of photons by tunneling electrons and on the base of perturbation theory calculations.\end{minipage}\vspace{2em}]

Physical phenomena in nanostructures caused by the influence of high-frequency electromagnetic fields on quantum electron transport have been considered in a large number of papers \cite{Dayem1962,Tien1963,Buttiker1982,Coon1985,Grifoni1998,Platero2004,Ge1996,Yakubo1996,Tkachenko1996,Morina2015,Kovalev2018}. First of all, resonant photon-stimulated tunneling was studied theoretically and experimentally in superconductor and semiconductor structures with atomically sharp potential barriers \cite{Dayem1962,Tien1963,Buttiker1982,Coon1985,Grifoni1998,Platero2004}. The creation of the quantum point contact (QPC), which is a lateral nanostructure consisting of a short, gate-driven constriction in a highly mobile two-dimensional electron gas (DEG) \cite{Wees1988,Wharam1988,Buttiker1990}, opened the way to studing photon-stimulated electron transmission through a smooth potential barrier, but until recently these studies were only theoretical \cite{Ge1996,Yakubo1996,Tkachenko1996}.

The authors of \cite{Otteneder2018,Tkachenko2018,Tkachenko2021} have discovered and studied for the first time the effect of the giant photoconductance of GaAs QPC under irradiation by terahertz radiation with photon energy $\hbar\omega_0=2.85$~meV, close to the difference between the Fermi energy and the top of the potential barrier $\hbar\omega_0=U_0-E_F$ (Fig. 1a). The effect was explained by the photon-stimulated transport (PST) of electrons due to the absorption of photons \cite{Otteneder2018,Tkachenko2018,Tkachenko2021}. However, the observed disappearance of the photoconductance  for a higher photon energy $\hbar\omega_0=6.74$~meV \cite{Otteneder2018}, has not received a clear qualitative explanation, although it agrees with the results of the numerical calculations \cite{Ge1996,Tkachenko2018}. Another (different from PST) picture of the formation of the giant photoresponse of QPC, based on the effect of modulation of the tunneling barrier height by the electromagnetic field, has been developed recently in \cite{Otteneder2021} to explain a strong superlinear dependence of the photoconductivity on the intensity of terahertz radiation in the deep tunneling mode. However, within the physical picture built in~\cite{Otteneder2021}, the question about the reason for the disappearance of photoconductivity when increasing frequency of radiation also remained open. In this paper, we propose an explanation of the disappearance of photoconductivity of QPC at high frequencies, based on the  quasimomentum conservation at optical transitions and justified by the calculation of PST spectra using the perturbation theory.
 
On a qualitative level, the explanation of the PST maximum at the photon energy close to the difference between the  top of the barrier and the Fermi level is as follows. In the presence of an electromagnetic wave with a frequency $\omega$, the energy conservation law allows electrons to undergo transitions to the Floquet states with energies $E_0\pm n\cdot\hbar\omega$, where $E_0$ is the initial energy, $n=1,2,\dots$ -- corresponds to the number of absorbed (+) or emitted (--) radiation quanta. Due to the momentum conservation law, the absorption of photons should occur with simultaneous scattering in momentum: by phonons or impurities in the bulk of the crystal, or by interacting with nanostructures. For nanostructures with quantum levels or quasi-levels in the absence of radiation, PST resonances usually appear as photon replicas of resonances in electron transport.

In a QPC with a single smooth barrier, in which there are no abrupt potential jumps, levels, or quasi-levels, there are no resonances in the energy dependence of the transmission coefficient $D(E)$ in the absence of radiation; as a consequence, such replicas cannot be observed. Nevertheless, in the photoconductance of such a QPC, a spectral resonance can be observed under the electron transitions to the top of the barrier. The reason for the resonance is illustrated in the energy-momentum diagram (Fig. 1b), which shows the electron dispersion laws near the stopping point and near the top of the barrier, as well as optical transitions with photon energies $\hbar\omega_0$ and $\hbar\omega_1$. It can be seen that for the ``resonant'' photon energy $\hbar\omega_0=U_0-E_F$, the optical transition from the bottom of the lower parabola to the bottom of the upper parabola is vertical and does not require additional scattering in momentum; therefore, the probability of such a transition is high. On the contrary, for $\hbar\omega_1>\hbar\omega_0$, the transition to a state with a high kinetic energy of an electron over the top of the barrier requires simultaneous scattering in momentum, so the probability of such a transition is small due to the small probability of acquiring a large momentum under transfer through a smooth barrier. Therefore, for $\hbar\omega>\hbar\omega_0$, the effect of PST decreases with the increase in $\hbar\omega$ due to the decrease in the photon absorption probability. For $\hbar\omega<\hbar\omega_0$, when the electron final energy is below the top of the barrier, PST increases with increasing photon energy due to the increase in the electron tunneling probability through the barrier. As a result, the magnitude of PST reaches its maximum at $\hbar\omega\approx\hbar\omega_0$. A similar picture of the formation of spectral maxima is also valid for multiphoton transitions with $n>1$.

\begin{figure}
\centering{\includegraphics*[width=1.0\linewidth]{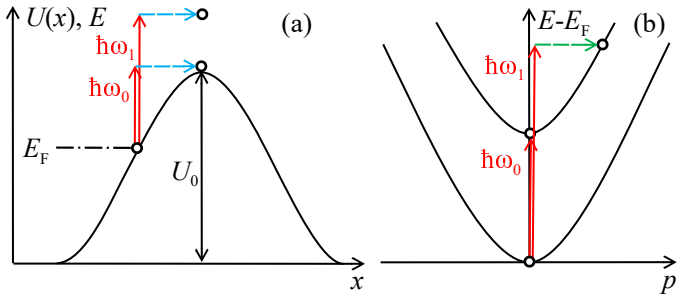}}

\caption{Illustration of the PST through a smooth potential barrier in the energy-coordinate (a) and energy-quasimomentum (b) diagrams when electrons absorb photons with energy $\hbar\omega_0=U_0-E_F$, corresponding to the transition to the top of the barrier, and for a higher photon energy $\hbar\omega_1>\hbar\omega_0$. In Fig. 1 (b) the lower and upper parabolas correspond to the dispersion laws of electrons near the stopping point and near the top of the barrier, respectively; required momentum scattering is indicated by a horizontal dashed arrow.}
\end{figure}

It is worth noting that in the proposed simplified picture, the ``resonant'' optical transition is direct (vertical) in $k$-space (Fig. 1b), but it is indirect in the real, $x$-space. That is, under the transition, the electron shifts from the stopping point to the top of the barrier (Fig. 1a). In reality, the stationary electron states are delocalized both in real space and in momentum space, so the simple picture of optical transitions between the states with well-defined dispersion laws is valid apparently only for sufficiently smooth barriers, for which the momentum uncertainty is relatively small. From the Heisenberg uncertainty relation, it is possible to estimate quasimomentum $\Delta p$ acquired by the electron under transfer through QPC and the half-width of the spectral maximum $\Delta E\sim(\Delta p)^2/2m^*$, where $m^*$ is the electron effective mass. Taking the uncertainty of the coordinate equal to the half-width of the potential barrier at the Fermi level $\Delta x\approx30$~nm, we obtain $\Delta E\approx0.5$~meV, which agrees by the order of magnitude with the results of the numerical calculations \cite{Tkachenko2021}.

Despite the simplicity and clarity of the proposed explanation of the ``resonance'' spectral maximum, the question remains open: To what extent this explanation is applicable to the real picture of the formation of the QPC photoresponse spectrum? Indeed, on the one hand, according to this explanation, due to the approximate equality of the wave vectors of the initial and final states in the spatial regions near the stopping points, these regions should give the main contribution to the transition matrix element. On the other hand, both the initial and final electron states belong to a continuous spectrum and are delocalized, therefore a broad region far from the stopping points can give a significant contribution to the transition matrix element. However, this contribution is weakened by the difference in the quasiclassical wave vectors; in particular, at a large distance from the potential barrier, optical transitions are not possible at all, since, as is known, a free electron cannot absorb a photon.

In order to find out which spatial region gives the main contribution to the spectral peak of the QPC photoresponse and to analyze the contribution of various factors to the formation of this peak, we have calculated PST spectra as the product of the optical transition probability $W$ and the electron transfer probability $D$ through the potential barrier in the final state. It should be noted that the idea to use such product in order to explain spectral maximum in PST was proposed, although not realized in~\cite{Ge1996}. We calculated the probability of the optical transitions from the initial state with Fermi energy $E_i=E_F$ to the final state with energy $E_f=E_F+\hbar\omega$ according to the first order perturbation theory, using the golden Fermi rule $W=2\pi/\hbar\cdot|<\psi_f|H'|\psi_i>|^2\delta(E_f-E_i-\hbar\omega)$. The wave functions of the initial $\psi_i$ and final $\psi_f$ states were taken from the solution of the problem of electron transfer through a smooth Eckart barrier with characteristic width $d$: $U(x)=U_0/cosh^2(x/d)$ \cite{Landau1977}. The wave functions were normalized to the electron flux $10^{12}$~c$^{-1}$. The upper part of Fig. 2 shows the plots of the squared modulus of the wave functions $|\psi(x)|^2$ in coordinate representation for three different electron energies. The bottom part of the figure shows the potential barrier $U(x)$, and the horizontal lines indicate the energies of the electrons. It is seen that for the low energy of the incident electron $E=25$~meV (5 meV below the top of the barrier), a standing wave is formed to the left of the barrier due to almost total reflection. For an electron energy equal to the barrier height $E=30$~meV, the reflection and transmission coefficients are approximately equal to 0.5, so there is a noticeable amplitude of the transmitted wave to the right of the barrier, while to the left the $|\psi(x)|^2$ value does not reach zero (it is about 3\% of the maximum). Finally, for energy $E=35$~meV (5~meV above the top of the barrier) there is almost complete transmission $D\approx0.99999$, the oscillation amplitude $|\psi(x)|^2$ associated with the over-barrier reflection is less than 1\%, and the $|\psi(x)|^2$ maximum is associated with quasiclassical deceleration, that is, with a decrease in electron velocity as it moves above the top of the barrier.

\begin{figure}
\centering{\includegraphics*[width=1.0\linewidth]{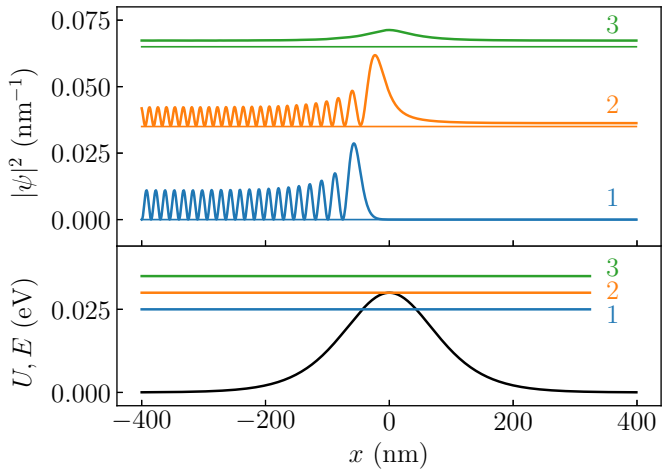}}

\caption{Top panel: squared wave function modulus $|\psi(x)|^2$ in coordinate space for the energies of incident electrons below the top of the barrier $E=25$~meV (1), equal with the top $E=30$~meV (2) and above the top $E=35$~meV (3). The plots of $|\psi(x)|^2$ for different energies are shifted in vertical direction for reader's convenience. Bottom panel: the Eckart potential barrier of height $U_0=30$~meV and width $d=100$~nm. The horizontal lines show the corresponding energies of the incident electrons.}
\end{figure} 

The Hamiltonian $H'$ of the interaction between electron and electromagnetic radiation was taken from~\cite{Anselm1981}, and the electron transmission coefficient for the final state $D(E_F+\hbar\omega_0)$ was taken from ~\cite{Landau1977}. The calculations were done for the intensity of terahertz radiation independent from $\hbar\omega$ and equal to 200~mW/cm$^2$. The one-dimensional density of states $\rho(E)=\sqrt{2m}/\pi\hbar\sqrt{E}$ was used for the kinetic energy of electrons $E=E_F+\hbar\omega$ far away from the barrier.

\begin{figure}
\centering{\includegraphics*[width=1.0\linewidth]{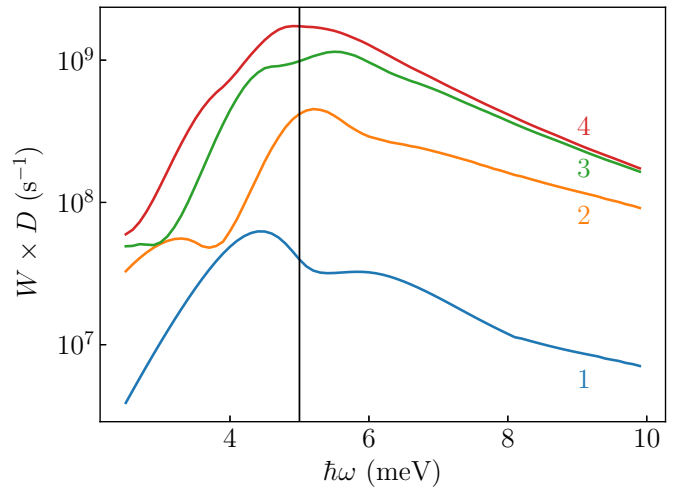}}
	
\caption{Photoresponse spectra calculated for QPC with barrier width $d=100$~nm, height $U_0=30$~meV, Fermi energy of electrons $E_F=25$~meV. The matrix element of optical transitions was integrated over  space region $|x|<L$, with various values of $L$ with respect to the characteristic barrier width $d$: 1 -- $L=d$; 2 -- $2d$; 3 -- $3d$; 4 -- $4d$. The ``resonant'' photon energy $\hbar\omega_0=U_0-E_F=5$~meV is indicated by the vertical line.}
\end{figure}

Fig. 3 shows the spectra of the PST magnitude $\approx W\times D$ calculated for a QPC with potential barrier width and height $d=100$~nm and $U_0=30$~meV, fixed energy $E_F=25$~meV and different $L$ values, where $L$ is the half-width of the space region from $-L$ to $+L$ around the barrier center, over which the matrix element of the optical transition was integrated. The spectral region is restricted by the photon energies $\hbar\omega>(U_0-E_F)/2$, because, for lower photon energies multi-photon processes, which are not accounted for in the present perturbation theory calculations, yield significant contribution to the photon-stimulated transport \cite{Tkachenko2018,Tkachenko2021}. It is seen that all spectra contain the leading peak centered approximately at the photon energy $\hbar\omega_0\approx U_0-E_F=5$~meV corresponding to optical transitions from the Fermi level to the top of the barrier. At $\hbar\omega<\hbar\omega_0$, the increase of the PST with increasing photon energy is due to the increase in transmission coefficient $D$; at $\hbar\omega>\hbar\omega_0$, $D$ saturates, and the decrease of the PST is due to the decrease in the probability of optical transition $W$; thus, the ``resonance'' peak is formed at $\hbar\omega\approx\hbar\omega_0$. These considerations are in line with a qualitative explanation of the peak in the QPC photoresponse spectum.

The comparison of the spectra for various values of $L$  shows that the main contribution to the resonance PST peak comes from the integration over the region $L\leq3d$; in this region the electron wave function is modified by the potential barrier so that the optical transitions become possible. It is seen that the decrease in the width of the integration region down to $L=d$ leads to drastic decrease in the peak amplitude, while for $L>3d$, amplitude and shape of the peak saturate with further increase in $L$. It should be noted that for the chosen energies and barrier parameters, the stopping point lie within $L\sim d$. This fact limits the applicability of the qualitative explanation proposed above, which is based on the assumption that the main contribution to the optical transitions comes from the regions near the stopping points. In fact, a significant contribution to the matrix element is given by a wider region of the potential barrier, including its ``foot''.

Note that relatively weak additional ``shoulders'' and extremes, which are superimposed on the leading peak for small integration regions $L$, are the ``side lobes'', arising due to the finite size of the integration window. Another artifact consisted of noise-like undulations that appeared in the calculated PST spectra due to variations in the phase of the integrand of the matrix element at the boundaries when $\hbar\omega$ changed; these oscillations were suppressed by averaging over the phase.

\begin{figure}
\centering{\includegraphics*[width=1.0\linewidth]{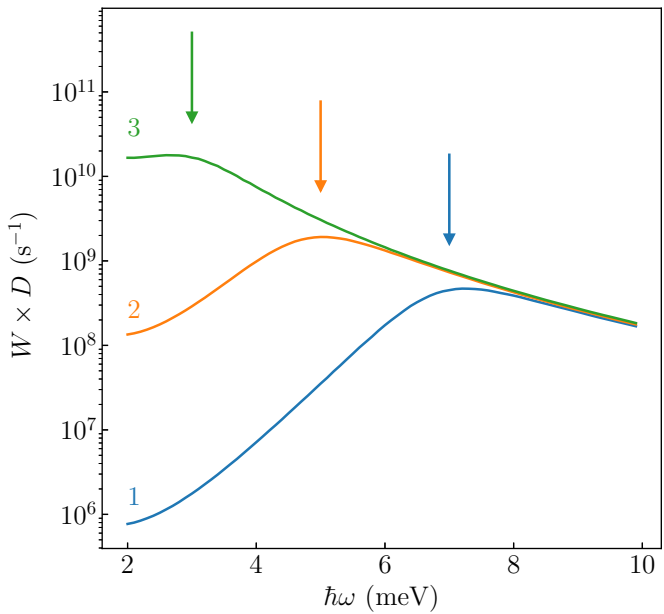}}
	
\caption{Calculated PST spectra  of a QPC for potential barrier width $d=100$~nm, height $U_0=30$~meV and various Fermi level positions: 1: $E_F=23$~meV; 2: 25~meV; 3: 27~meV. For each spectrum, the respective ``resonant'' photon energy $\hbar\omega_0=U_0-E_F$ is indicated by the arrow.}
\end{figure}

Fig. 4 shows the PST spectra calculated for the fixed height of the potential barrier $U_0$ and various Fermi levels $E_F$. It is seen that for all values of $E_F$, the spectral position of the leading peak corresponds to the optical transitions from the Fermi level to the top of the potential barrier, in accordance with the results of the numerical calculations \cite{Tkachenko2021} and with the qualitative explanation of PST proposed here (Fig. 1). With increasing $E_F$, the peak broadens and becomes less distinct. It is also seen that the slope of the high-energy tail of the peak weakly depends on the Fermi level. 

\begin{figure}
\centering{\includegraphics*[width=1.0\linewidth]{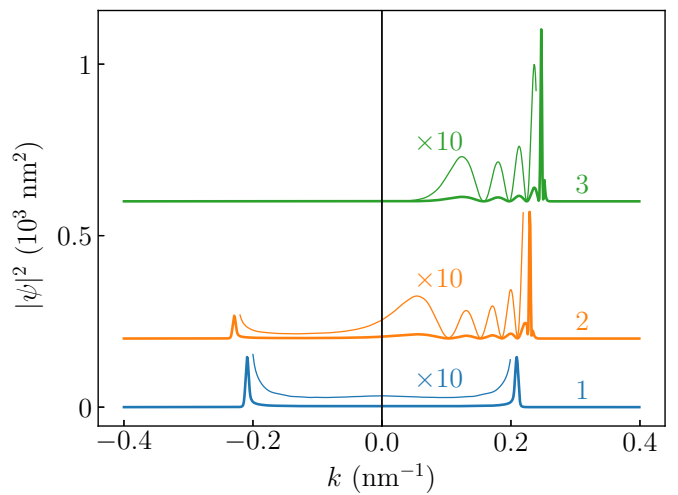}}
	
\caption{Squared wave function modulus $|\psi(k)|^2$ in momentum space for energies of incident electrons below ($E=25$~meV, 1), equal ($E=30$~meV, 2) and above ($E=35$~meV, 3) the barrier top . For reader's convenience, the parts of the wave functions between the delta-shaped peaks corresponding to the incident and reflected electrons are magnified by a factor of 10, and the plots for the different energies are shifted in the vertical direction.}
\end{figure} 

To further clarify the question of the maximum formation mechanism, we calculated the PST spectrum using the wave functions of the electrons in the momentum space. The squared modulus of these functions $|\psi(k)|^2$ for different energies of the incident electrons are shown in Fig. 5. It can be seen that for an electron energy 5 meV below the barrier height, $|\psi(k)|^2$ consists of two delta-like peaks corresponding to the incident (right peak) and reflected (left peak) electron momenta, and a weak symmetric monotonic ``background'' between the peaks, which reflects the deceleration of the incident and acceleration of the reflected electron on the left slope of the potential barrier. For an energy equal to the height of the barrier, the transmission is $D\sim0.5$; accordingly, the left delta peak is significantly smaller than the right one. In this case, the wave function $|\psi(k)|^2$ between delta peaks becomes asymmetric: the region of positive momenta dominates, in which additional relatively broad peaks are observed. For energies 5 meV above the barrier height, the left delta peak is not observed, since there is almost no reflected wave, and the additional peaks are shifted toward larger positive momenta. The position of the most pronounced peak, which lies closer to $k=0$, corresponds to the momentum of electrons moving in the region near the top of the barrier. Additional peaks at higher momenta are probably ``side lobes'' (``Gibbs oscillations'') of $|\psi(k)|^2$, arising in the Fourier transform due to localization of the electron wave function in the barrier region (Fig. 2).

The momentum space wavefunction $|\psi(k)|^2$ (Fig. 5) allows us to give an additional explanation of the role of the momentum conservation law in the effect of photoconductivity disappearance at high photon energies. Indeed, in the initial state (the lower curve in Fig. 5) the average momentum is zero, while in the final state above the barrier top $E>U_0$ the momentum is different from zero and corresponds to almost complete transmission of the electron. As a consequence, the probability of an optical transition between these states is small and decreases with increasing photon energy due to the momentum conservation law.

\begin{figure}
\centering{\includegraphics*[width=1.0\linewidth]{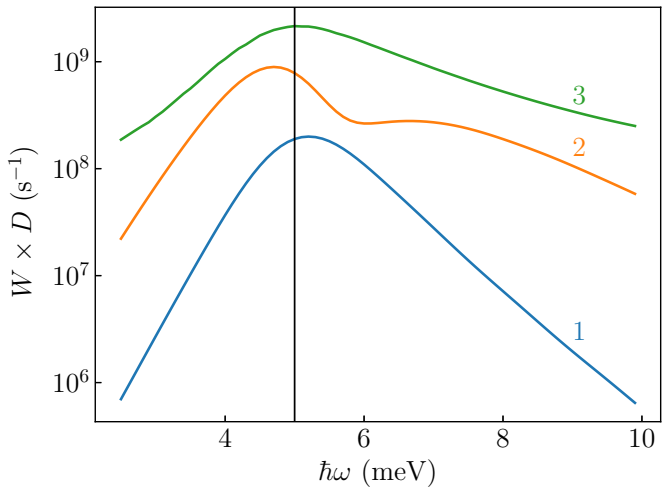}}
	
\caption{Photoresponse spectra calculated for QPC with barrier width $d=100$~nm, height $U_0=30$~meV, Fermi energy of electrons $E_F=25$~meV. The matrix element of optical transitions was integrated over momentum space region $|k|<K$, with various values of $K$ (wwith respect to the wave vector of electrons before ($k_1$) and after ($k_2$) the optical transition): 1 -- $K=0.25k_1$; 2 -- $0.5k_1$; 3 -- $1.1k_2$. The ``resonant'' photon energy $\hbar\omega_0=U_0-E_F=5$~meV is indicated by the vertical line.}
\end{figure}

Fig. 6 shows PTS spectra calculated for a QPC with potential barrier width and height $d=100$~nm and $U_0=30$~meV, for a fixed Fermi energy $E_F=25$~meV and for various values of $K$, where $K$ is the half-width of a region in $k$-space from $-K$ to $+K$, over which the matrix element of the optical transition was integrated. It can be seen that, similarly to the calculation of the matrix element from the coordinate wave functions (Fig. 3), the main peak in all spectra is centered approximately near the photon energy $\hbar\omega_0=U_0-E_F=5$~meV, which corresponds to optical transitions from the Fermi level to the top of the barrier. Comparison of the spectra in Fig. 6 for various $K$ values shows that a significant contribution to the PST resonance peak is made by integrating over the entire region between the delta peaks corresponding to the incident and transmitted wave, and not just the region near the stopping points, where the quasiclassical momentum is close to zero. This is consistent with the conclusion drawn from Fig. 3 that a significant contribution to the matrix element of the optical transition is made by the ``foot'' of the barrier, and not only by the regions near the stopping points, as was assumed in the simple qualitative explanation of the origin of the resonance peak in the PST spectrum illustrated in Fig. 1. Nevertheless, we believe that this explanation correctly points to the main reason for the experimentally observed drop in the PST magnitude when the photon energy exceeds the resonance value. This reason consists in the drop in the matrix element of the optical transition at a large difference between the momenta of the initial and final states of the electron.

Considerations about momentum conservation during optical transitions are of a general nature and can be used to explain not only the PST spectra of QPC, but also to qualitatively interpret the photoionization spectra of other physical objects and, in particular, the hydrogen atom. The photoionisation cross section of a hydrogen atom decreases with increasing photon energy in all known models, including the Born approximation and the Sommerfeld model, which takes into account the Coulomb interaction of the electron with the ion~\cite{Astapenko2010}. It is useful to look at this decrease from the point of view of the momentum conservation law during an optical transition from a bound state to the continuous spectrum. Indeed, in a bound state, an electron does not have a definite momentum, but there is a momentum distribution, the width of which is about $\hbar/a_B$, where $a_B$ is the Bohr radius. One can say that a photon ``grabs'' an electron from this distribution and transfers it to the continuous spectrum. Neglecting the small momentum of the photon, the optical transition in the energy-momentum diagram can be considered vertical. The probability of finding an electron with a certain momentum in a bound state decreases with increasing momentum. As a consequence, as the photon energy increases, the photoionization cross section decreases. The characteristic width of the tail of the photoionization spectrum corresponds to the momentum of the ionized electron $k\sim\hbar/a_B$, in agreement with the above qualitative explanation. Similar considerations are valid for explaining the photoionization spectrum of shallow hydrogen-like impurity centers in semiconductors.

Thus, considerations about the momentum conservation upon absorption of photons provide a qualitative explanation for the nonmonotonic (``resonant'') dependence of the photoconductance of a QPC with a maximum near the photon energy $\hbar\omega_0=U_0-E_F$, which was observed experimentally \cite{Otteneder2018} and obtained by numerical calculations \cite{Tkachenko2021}. Within the framework of the proposed picture, the width of the spectral maxima has been estimated and the evolution of the spectral shape with changing the position of the Fermi level has been explained qualitatively. The calculation of the PST spectrum by perturbation theory is qualitatively consistent with the proposed explanation, but it imposes quantitative restrictions on the explanation.

This study was supported by the Ministry of Science and Higher Education of the Russian Federation (state assignment for the Rzhanov Institute of Semiconductor Physics, Siberian Branch, Russian Academy of Sciences).

\renewcommand\refname{{\normalsize References}}

\end{document}